\documentclass[letterpaper, paper,11pt]{IEEEtran}		

\usepackage{bm}
\usepackage{amsmath}
\usepackage{amsfonts}
\usepackage{subfigure}
\usepackage{graphicx}
\usepackage[colorlinks=true, pdfstartview=FitV, linkcolor=black, citecolor= black, urlcolor= black]{hyperref}
\usepackage{overcite}
\usepackage{footnpag}			      	

\begin{document}

\title{Catalog and Characterization of Science Orbit Configurations for an Enceladus Orbiter}

\author{Spencer Boone\thanks{Fédération ENAC ISAE-SUPAERO ONERA, 
Université de Toulouse, France, spencer.boone@isae-supaero.fr}, 
Joan Pau Sánchez Cuartielles\thanks{Fédération ENAC ISAE-SUPAERO ONERA, 
Université de Toulouse, France, joan-pau.sanchez@isae-supaero.fr},  
\ and Stéphanie Lizy-Destrez\thanks{Fédération ENAC ISAE-SUPAERO ONERA, 
Université de Toulouse, France, stephanie.lizy-destrez@isae-supaero.fr}
}

\maketitle{}

\begin{abstract}
Saturn's moon Enceladus is an exciting destination for future exploration missions due to the scientifically interesting geyser region located on its South pole. In this work, we compile the different types of science orbit configurations that have been proposed in the literature and present numerical methods to compute each of them in the Saturn-Enceladus circular restricted three-body problem (CR3BP). In addition, we explore the utility of the higher-period dynamical structures found in the CR3BP. Figures of merit such as the observational properties and geometries for each family of orbits are presented. By providing a consistent analysis of potential Enceladus science orbits, this work can serve as a baseline for future mission designs.
\end{abstract}

\section{Introduction}

\begin{figure}

        \centering
        \includegraphics[scale=0.2]{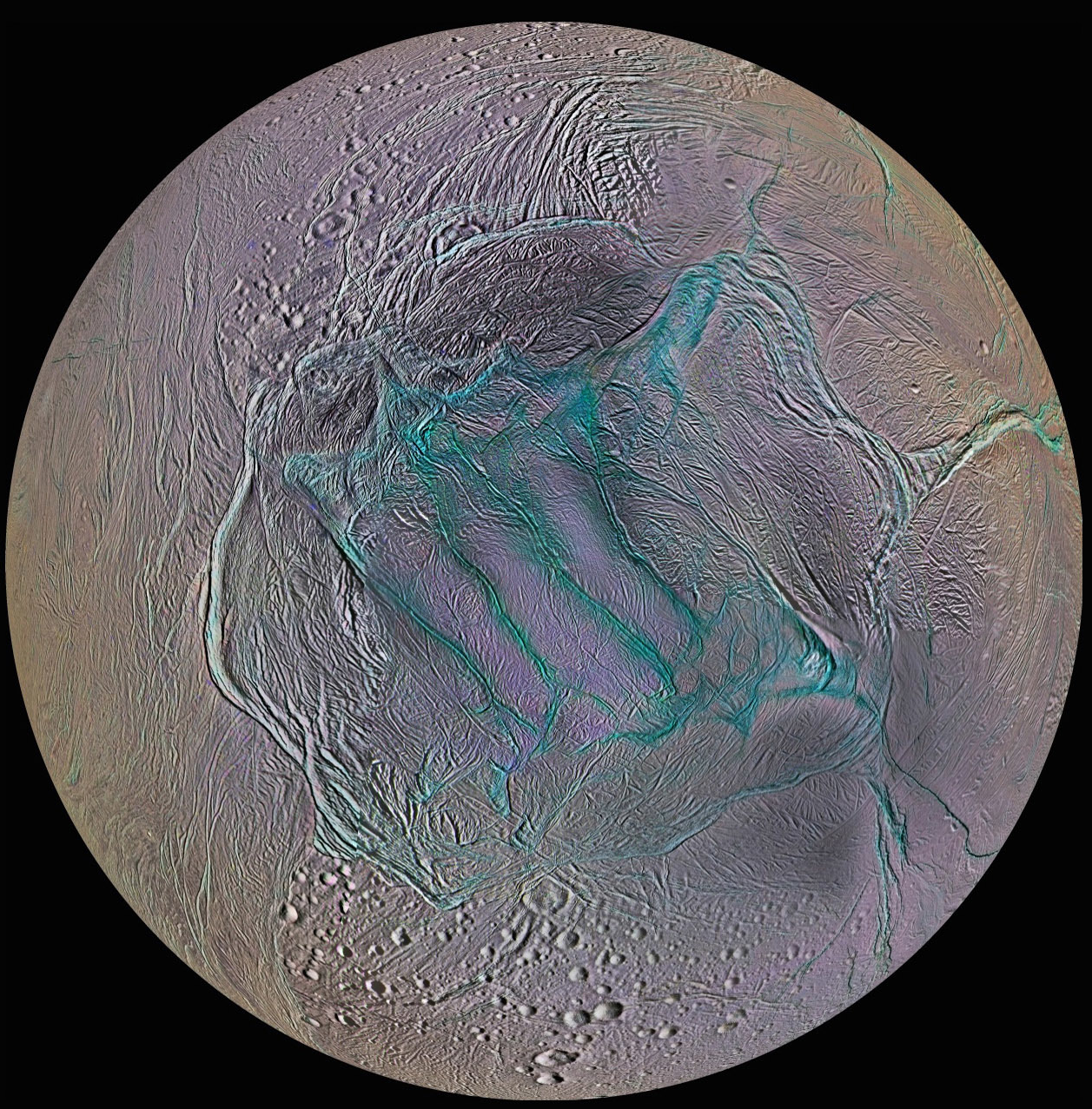} 
        \caption{Enhanced view of Enceladus' South pole region showing the locations of organic-rich geysers in blue [Source: NASA]}
        \label{f:enceladus_tiger_stripes}

\end{figure}

Saturn's moon Enceladus has represented an exciting destination for future exploration missions since the discovery of organic-rich geysers on its South pole region by the Cassini spacecraft. In particular, a spacecraft orbiting around or in the vicinity of Enceladus would be able to obtain valuable measurements to assess the moon's ability to sustain simple life forms. For mission designers, designing a science orbit in this regime represents a unique challenge, due to Enceladus' small size with respect to Saturn and the unique location of interest at the moon's South pole - see Fig.~\ref{f:enceladus_tiger_stripes}.

Accordingly, a number of studies have investigated science orbit configurations specifically for a future Enceladus mission~\cite{russell_lara, davis_enceladus_nrho, salazar_cr3bp_oblate, parker_enceladus}. However, with a couple exceptions, these studies suggest only a single orbit option and do not provide a detailed comparison or analysis of the tradeoffs for different types of orbit. Examples of proposed configurations include inclined near-circular orbits~\cite{russell_lara}, near rectilinear halo orbits (NRHOs)~\cite{davis_enceladus_nrho}, and heteroclinic connections between more distant halo orbits~\cite{salazar_cr3bp_oblate, davis_enceladus_nrho}. These orbits are shown to be useful, but each work does not use the same set of criteria to evaluate their suitability. These works also use different dynamical models to approximate the dynamics in the Enceladus-Saturn system, such as the Hill three body problem with oblate bodies~\cite{russell_lara}, the circular restricted three-body problem~\cite{davis_enceladus_nrho}, the circular restricted three-body problem with oblate bodies~\cite{salazar_cr3bp_oblate}, and a high-fidelity ephemeris model~\cite{parker_enceladus}. These discrepancies can make it difficult for a mission designer to efficiently compare these orbits during the preliminary mission design phase.

In this work, we investigate the range of science orbit configurations and assess their utility for achieving the science objectives for an Enceladus orbiting mission. For the most similar dynamical system that has been extensively studied in the literature (the Jupiter-Europa system) most work has focused on finding science orbits that achieve global mapping of Europa's surface. The design of a future Enceladus mission requires exploring various orbit configurations that can be used to accurately map the unique area of interest at the South Pole region.

In addition, we provide a thorough investigation of the periodic dynamical structures that exist in the context of the circular restricted three-body problem. In the current literature, mainly the halo orbit family~\cite{davis_enceladus_nrho} and the nearby ''period-3 halo orbits"~\cite{orbilander} have received considerable attention. However, additional families of periodic orbits bifurcate from the halo orbit family, which could have favorable properties to achieve variation in observation geometry over Enceladus' South pole region. Extensive analysis has been conducted to characterize these various families for the Earth-Moon~\cite{zimovan_nrhos_higher_period}, Jupiter-Europe~\cite{luke_europa}, and Jupiter-Callisto systems~\cite{yang_families_jupiter_callisto}. We conduct an analogous investigation of orbit families in the Saturn-Enceladus CR3BP.

This paper is organized as follows. We first present an analysis of the different dynamical models that are used in the literature to approximate the dynamics in the vicinity of Enceladus, and will motivate the selection of a model with an appropriate balance between fidelity and simplicity. Next, we will present the various types of orbits in this model, along with a description of the numerical tools used to compute them. First, we present the halo orbit families and nearby higher-period bifurcations from this family. Next, other potentially interesting families of orbits such as the axial and butterfly orbits are investigated. Finally, we compute families of low-energy heteroclinic connections between halo orbits that could provide unique viewing geometries. This work provides a thorough and consistent analysis of Enceladus science orbits that can serve as a baseline for future mission designs.

\section{Dynamic models}

When performing preliminary analysis of science orbits, it is important to select the appropriate dynamic model, balancing speed, ease of implementation, and fidelity. As stated in the introduction, a number of dynamic models have been used to generate science orbits in the Saturn-Enceladus system. Each of these models has varying levels of fidelity and implementation complexity, which motivates a comparison of these systems.

\subsection{Circular restricted three-body problem}
The most commonly-used system for modeling multibody dynamics is the circular restricted three-body problem (CR3BP). The equations of motion for the CR3BP are
\begin{align}
&\ddot{x} = 2\dot{y} + x - \frac{(1 - \mu)(x + \mu)}{r_1^3} - \frac{\mu(-1 + x + \mu)}{r_2^3}
\label{eq:crtbpx}\\
&\ddot{y}  = -2\dot{x} + y - \frac{y(1 - \mu)}{r_1^3} - \frac{\mu y}{r_2^3} 
\label{eq:crtbpy}\\
& \ddot{z} = -\frac{z(1-\mu)}{r_1^3} - \frac{\mu z}{r_2^3} \label{eq:crtbpz}
\end{align}
where the state vector contains three position $(x,y,z)$ and three velocity $(\dot{x}, \dot{y}, \dot{z})$ terms. In the CR3BP, $\mu$ is the ratio of the mass of the secondary body to the total system mass. The CR3BP relies on the assumption that the mass of the spacecraft is negligible compared to the masses of the primary and secondary bodies. The system dynamics are modeled in a rotating frame. The distances $r_1$ and $r_2$ are defined as the distances from the spacecraft to the primary and secondary bodies, respectively. Distance and time units are normalized with respect to the distance between the two bodies and their orbital period, respectively, which we will refer to as $LU$ and $TU$. For preliminary mission design in the Saturn-Enceladus system, several works in the literature have used the CR3BP~\cite{davis_enceladus_nrho, parker_enceladus, orbilander}.

\subsection{Hill three body problem}
A simplified form of the CR3BP when the mass ratio $\mu \rightarrow 0$ is the Hill three body problem (H3BP)~\cite{szebehely}. A series of papers by Lara and Russell~\cite{russell_lara, lara_russell_enceladus} investigate low-altitude science orbits in the context of the H3BP or the H3BP augmented with spherical harmonic terms. The H3BP may be less accurate than the CR3BP for orbits far from Enceladus' sphere of influence. In addition, with the increased number of missions operating or proposed to operate in the Earth-Moon system, a variety of numerical tools have been developed for computing orbits specifically in the Earth-Moon CR3BP. Because of this growing catalog of literature using the CR3BP, it was determined that the simplicity of the H3BP is not sufficient to justify using it over the CR3BP for this work.

\subsection{Circular restricted three-body problem with oblate primaries}
The circular restricted three-body problem can be extended to include the spherical harmonics of either (or both) the primary and secondary bodies, accounting for the gravitational effects of their non-spherical shape. Several works have considered this type of dynamic model: for example, in the work of Salazar et al.~\cite{salazar_cr3bp_oblate}, the Saturn-Enceladus CR3BP is augmented with the $J_2$ spherical harmonic terms for both Saturn and Enceladus. In Bury and McMahon.~\cite{luke_zonal_harmonics}, a set of recursive equations are derived to include any number of spherical harmonic terms in the CR3BP. 

For the Saturn-Enceladus system, the $J_2$ terms are by far the largest of these terms, so we consider only the dynamic system including these. We use the equations derived in Salazar et al.~\cite{salazar_cr3bp_oblate}, with minor modifications required in order to use the alternative convention of placing the primary and secondary at $[-\mu, 0, 0]$ and $[1-\mu, 0, 0]$ respectively, rather than at $[\mu, 0, 0]$ and $[\mu - 1, 0, 0]$.

The equations of motion for the CR3BP with primary and secondary $J_2$ effects are thus as follows. First, one must normalize the units of the dynamic system as in the CR3BP. The normalized spherical harmonic parameters thus become $A_1 = \frac{J_{2,1} R_1^2}{LU^2}$ and $A_2 = \frac{J_{2,2} R_2^2}{LU^2}$, where $J_{2,1}$ and $J_{2,2}$ correspond to the $J_2$ parameters for the primary and secondary bodies, respectively. The normalized mean motion of the system can thus be defined as
\begin{equation}
    \hat{n} = \sqrt{1 + \frac{3(A_1 + A_2)}{2}}
\end{equation}

The equations of motion thus become 
\begin{align}
    & \ddot{x} = 2 n \dot{y} + n^2 x - \frac{(1 - \mu)(x + \mu)}{r_1^3} C_1 - \frac{\mu(-1 + x + \mu)}{r_2^3} C_2
\label{eq:crtbp_oblate_x}\\
&\ddot{y}  = - 2 n \dot{x} + n^2 y - \frac{y(1 - \mu)}{r_1^3} C_1 - \frac{\mu y}{r_2^3} C_2
\label{eq:crtbp_oblate_y}\\
& \ddot{z} = -\frac{z(1-\mu)}{r_1^3} \tilde{C}_1 - \frac{\mu z}{r_2^3} \tilde{C}_2  \label{eq:crtbp_oblate_z}
\end{align}
where
\begin{align}
& C_i = 1 - \frac{3}{2} \frac{A_i}{r_i^2} \left[ 5 \left( \frac{z}{r_i} \right)^2 - 1 \right], \quad  i = 1,2
\label{eq:crtbp_oblate_C1}\\
& \tilde{C}_i = C_i + 3 \frac{A_i}{r_i^2}, \quad  i = 1,2
\label{eq:crtbp_oblate_C2}
\end{align}
Equivalent forms for the effective pseudo-potential and Jacobi constants in this perturbed system are provided in the literature~\cite{salazar_cr3bp_oblate, luke_zonal_harmonics}. Equations to numerically integrate the state transition matrix in this system are also given available~\cite{salazar_cr3bp_oblate}, though we note again that this reference uses alternative locations for the primary and secondary bodies, which affects the signs of some terms in these expressions.

\subsection{Ephemeris model}
For detailed analysis of science orbits around Enceladus, a full-fidelity model of the dynamics can become useful. The Sun, all planets in the solar system and all of Saturn's moons can be included in this model, along with solar radiation pressure and spherical harmonic models for Saturn and Enceladus. This model provides the highest fidelity of dynamics, but may be too computationally expensive for the purposes of preliminary orbit design. Davis et al.~\cite{davis_enceladus_nrho} provide examples of correcting orbits initially computed in the Saturn-Enceladus CR3BP into an ephemeris model, and the science orbit analyses done by Parker et al.~\cite{parker_enceladus} appear to have been computed in a higher-fidelity model. In this work, because much of the preliminary science orbits consider structures in the three-body problem, we do not investigate the use of an ephemeris model.

\subsection{Evaluation of dynamic models}
\begin{figure}
    \centering

        \centering
        \includegraphics[scale=0.6]{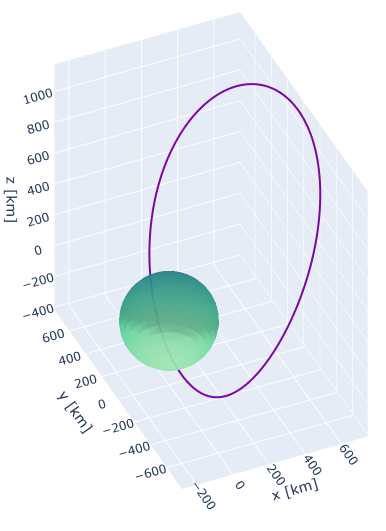} 
        \caption{$L_2$ NRHO computed in the oblate CR3BP}
        \label{f:ocr3bp_nrho}

        \centering
        \includegraphics[scale=0.65]{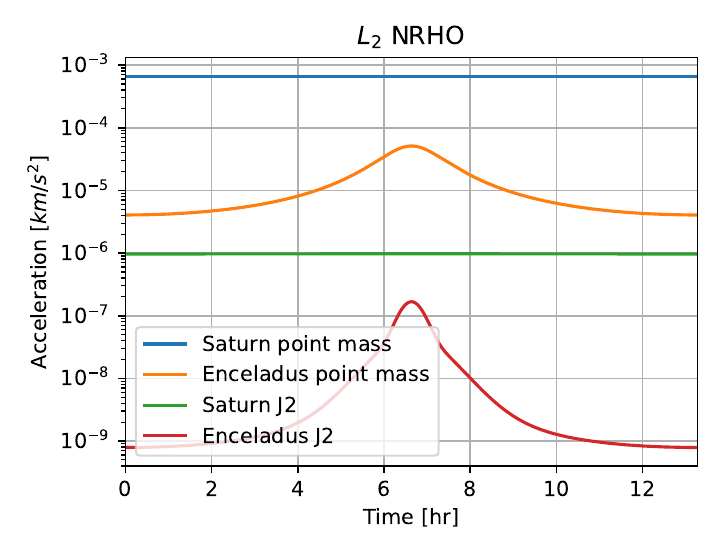} 
        \caption{Acceleration magnitudes for $L_2$ NRHO integrated over one period}
        \label{f:ocr3bp_nrho_accs}

\end{figure}

In order to understand the effects of the Saturn and Enceladus spherical harmonics on the evolution of science orbits near Enceladus, we present an example of a potential science orbit computed in the oblate CR3BP. The parameters for this system are given in Table~\ref{tab:params}, obtained from JPL SPICE toolkit~\cite{jpl_spice}. Using ISAE-SUPAERO's SEMPy toolbox~\cite{sempy}, an $L_2$ near rectilinear halo orbit was computed in the Saturn-Enceladus oblate CR3BP. This orbit is shown in Fig.~\ref{f:ocr3bp_nrho}. The order of magnitude of each of the accelerations in this system were computed over one period of the orbit and are shown in Fig.~\ref{f:ocr3bp_nrho_accs}. The effects of Saturn's oblateness are clearly more important than Enceladus', even when the spacecraft during the close periapse passage. In addition, the acceleration due to Enceladus' gravity is always more important than that due to Saturn's oblateness. These findings align with the analysis conducted by Salazar et al.~\cite{salazar_cr3bp_oblate}. Due to this limited importance of the oblateness terms for preliminary mission design, we chose to conduct the remainder of our analysis using the standard CR3BP dynamic model.

\begin{table}[b]
\centering
\caption{Saturn-Enceladus system parameters}
\begin{tabular}{cc}
\hline \hline
Parameter & Value\\
\hline
$\mu$ & $1.9011497893288988 \times 10^{-7}$\\
$R_{E}$ & 252.1 km\\
$R_{S}$ & 58300.0 km\\
$J_{2,E}$ & 0.0025\\ 
$J_{2,S}$ & 0.016298\\
$LU$ & 238400 km\\
$TU$ & 18899.965056082812 s\\
\hline \hline
\label{tab:params}
\end{tabular}
\end{table}

\section{Halo orbits and associated higher-period bifurcations}
The first type of science orbit investigated in this work is the halo orbit family of periodic orbits. This family of orbits is potentially very useful due to their ability to obtain repeated close passages over the scientifically interesting South pole region of Enceladus. Higher-period dynamical structures near the halo orbit families can also be generated, which may have favorable observational properties, as shown in previous work\cite{orbilander,zimovan_nrhos_higher_period}.

\subsection{$L_1$ and $L_2$ halo orbit families}
\begin{figure}
    \centering

        \centering
        \includegraphics[scale=0.5]{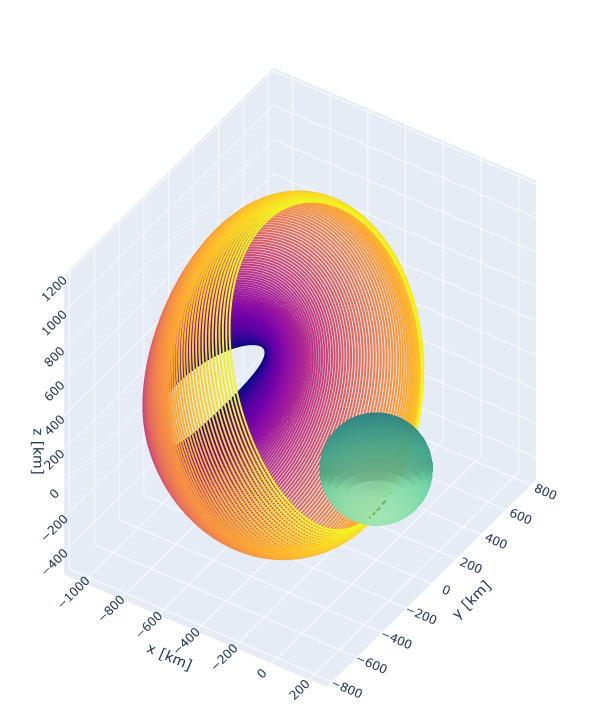} 
        \caption{$L_1$ family of halo orbits}
        \label{f:l1_halo_orbits}

        \centering
        \includegraphics[scale=0.65]{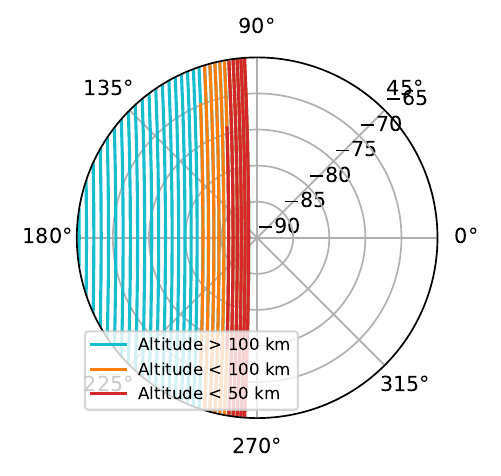} 
        \caption{$L_1$ halo orbit groundtracks over South pole}
        \label{f:l1_halo_orbit_gts}
        \includegraphics[scale=0.6]{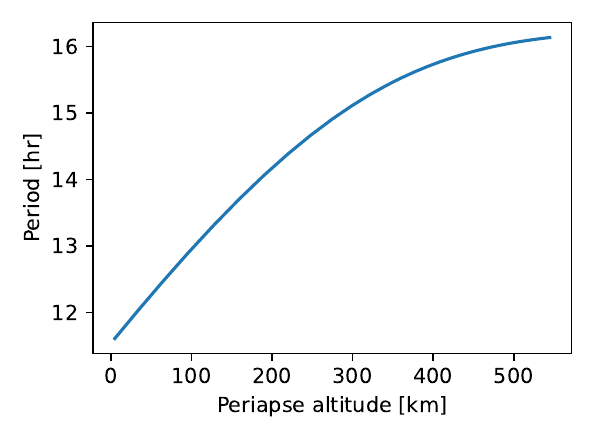} 
        \caption{$L_1$ halo orbit periods vs. periapse altitude}
        \label{f:l1_halo_orbit_period_alt}

    \end{figure}

\begin{figure}

    \centering

        \centering
        \includegraphics[scale=0.5]{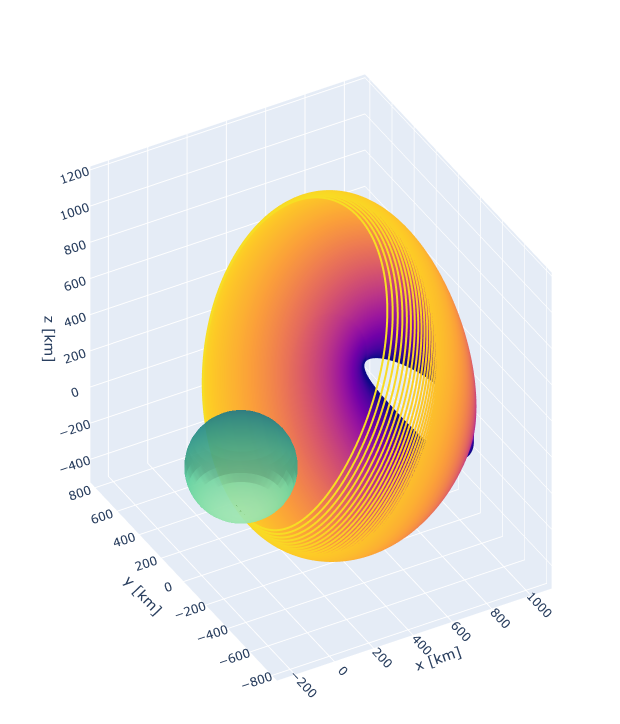} 
        \caption{$L_2$ family of halo orbits}
        \label{f:l2_halo_orbits}

        \centering
        \includegraphics[scale=0.65]{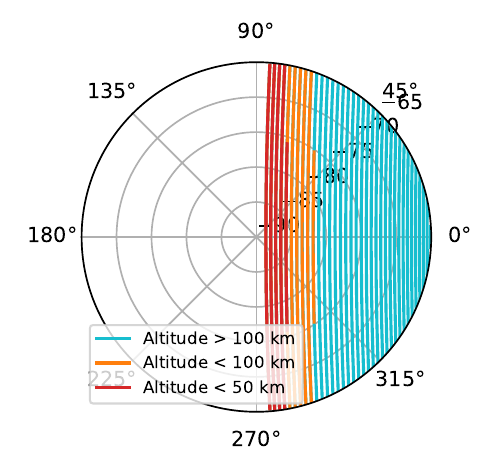} 
        \caption{$L_2$ halo orbit groundtracks over South pole}
        \label{f:l2_halo_orbit_gts}
        \includegraphics[scale=0.6]{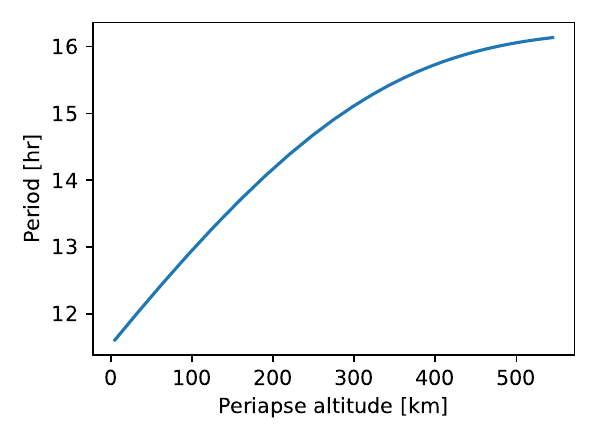} 
        \caption{$L_2$ halo orbit periods vs. periapse altitude}
        \label{f:l2_halo_orbit_period_alt}
    
\end{figure}

We can first analyze the properties of the halo orbit families about the $L_1$ and $L_2$ Lagrange points in the Saturn-Enceladus CR3BP. Using the open-source SEMPy software tools developed at ISAE-SUPAERO~\cite{sempy}, the $L_1$ and $L_2$ halo orbit families were continued beginning from their bifurcation from the $L_1$ and $L_2$ Lyapunov orbit families with the higher-order polynomial continuation method~\cite{andrea_continuation}. The subset of these families that do not impact into the surface of Enceladus are shown in Figs.~\ref{f:l1_halo_orbits} and~\ref{f:l2_halo_orbits}. 

The groundtracks for the halo orbit families are shown in Figs.~\ref{f:l1_halo_orbit_gts} and~\ref{f:l2_halo_orbit_gts}. The periods of each orbit within the families are shown as a function of their periapse altitude in Figs.~\ref{f:l1_halo_orbit_period_alt} and~\ref{f:l2_halo_orbit_period_alt}. The periods of these halo orbits vary from around 11 to 16 hours, meaning they can offer frequent passes over the South pole region of the moon. However, this may introduce operational complexities depending on the stationkeeping requirements, since it would not be operationally feasible to plan maneuvers on the ground and execute them every orbit. These orbits possess favorable groundtracks over Enceladus' South pole region, with the lowest-altitude members of the family possessing particularly interesting close passages over Enceladus' South pole region. This subset of the family is referred to as the near rectilinear halo orbits. These have been proposed as candidates for Enceladus science orbits in several references~\cite{davis_enceladus_nrho, orbilander, esa_enceladus}.


\subsection{Higher-period halo orbit family bifurcations}
A potential downside of selecting a member of the NRHO family as a science orbit is the lack of variation in groundtrack coverage between orbits. Because Enceladus is tidally locked with Saturn, the NRHOs will perfectly repeat their groundtracks at every passage of the South pole (in the CR3BP), and this behaviour will also be preserved when transitioned to an ephemeris model. This motivates investigating the higher-period bifurcations from the halo orbit family, in order to identify whether any of these more complex structures have favorable groundtrack properties with more variation in location. For example, a period-3 halo orbit was selected as the science orbit for the proposed Enceladus Orbilander mission~\cite{orbilander}.

\begin{figure}
\begin{center}
\includegraphics[scale=0.7]{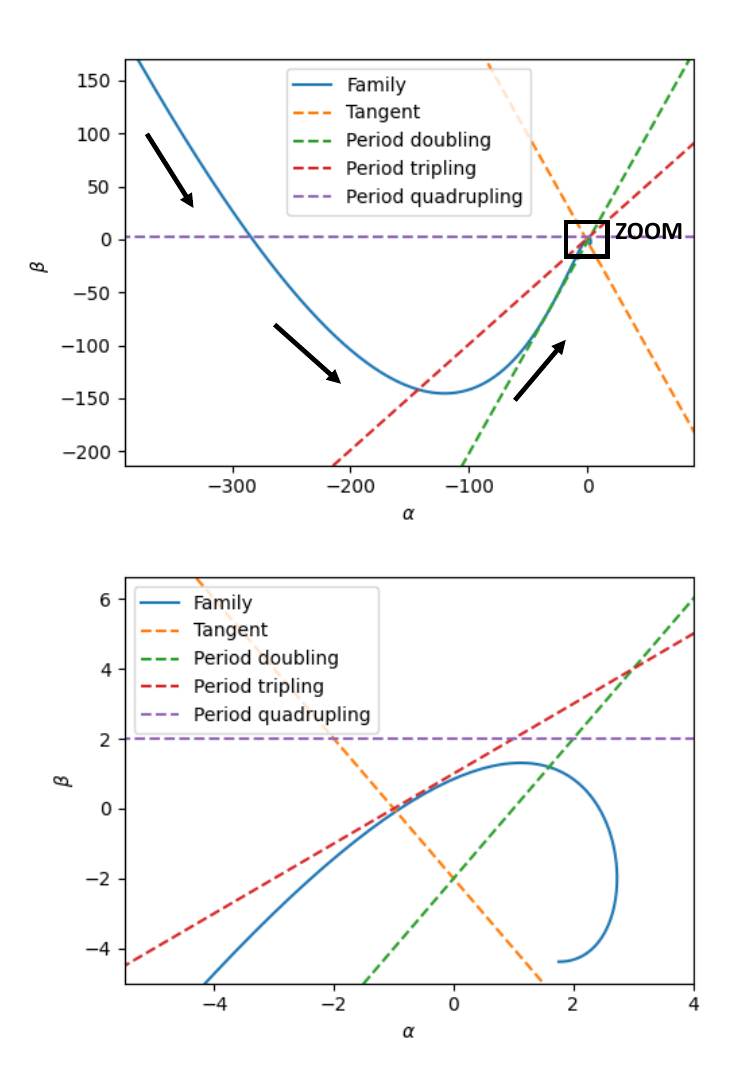}
\caption{Broucke stability diagram for $L_2$ halo orbit family in Saturn-Enceladus system}
\label{f:broucke}
\end{center}
\end{figure}

A similar analysis was previously conducted by Zimovan-Spreen and Howell~\cite{zimovan_nrhos_higher_period} for the Earth-Moon $L_2$ halo orbit family. We follow the methodology described in this work to identify the higher-period bifurcations for the Saturn-Enceladus halo orbit families. First, the \emph{Broucke stability diagram}~\cite{broucke_stab} is generated for both the $L_1$ and $L_2$ halo orbit families. In the Broucke stability diagram, two scalars $\alpha$ and $\beta$ are introduced that fully define the nontrivial eigenvalues of the monodromy matrix of each orbit. These scalars are plotted against each other, resulting in a curve. When the curve passes one of the bifurcation lines indicated on the stability diagram, a bifurcation of the associated type occurs, and a new family orbits can be computed and continued. Further details on the computation of the scalars and the definitions of the bifurcation lines can be found in Zimovan-Spreen and Howell~\cite{zimovan_nrhos_higher_period}. For reference, the Broucke stability diagram for the $L_2$ halo orbit family is shown in Fig.~\ref{f:broucke}.

We use the SEMPy toolbox~\cite{sempy} to automatically identify these bifurcations and continue the resulting families. A total of three period-doubling and one period-tripling bifurcation are identified for both the $L_1$ and $L_2$ halo orbit families - these can be seen for the $L_2$ halo orbit family in Fig.~\ref{f:broucke}. New families of periodic orbits can then be continued starting from the bifurcation points. An interesting observation is that the fourth period-doubling bifurcation that is observed for the halo orbit family in the Earth-Moon system~\cite{zimovan_nrhos_higher_period} does not appear on the Broucke stability diagram for the Saturn-Enceladus system, even when the family is continued very close to the center of Enceladus. When the new family of orbits is continued from this fourth bifurcation, we obtain what are commonly known as the \emph{butterfly} orbits~\cite{zimovan_nrho_butterfly}. These orbits are known to exist in the Saturn-Enceladus system~\cite{jpl_database}, but their bifurcation from the halo orbit family appears to exist inside a singularity. They therefore need to be generated either from known initial conditions or by continuing along the mass ratio $\mu$ from a system with larger $\mu$ where this bifurcation point exists.

\begin{figure}

        \centering
        \includegraphics[scale=0.5]{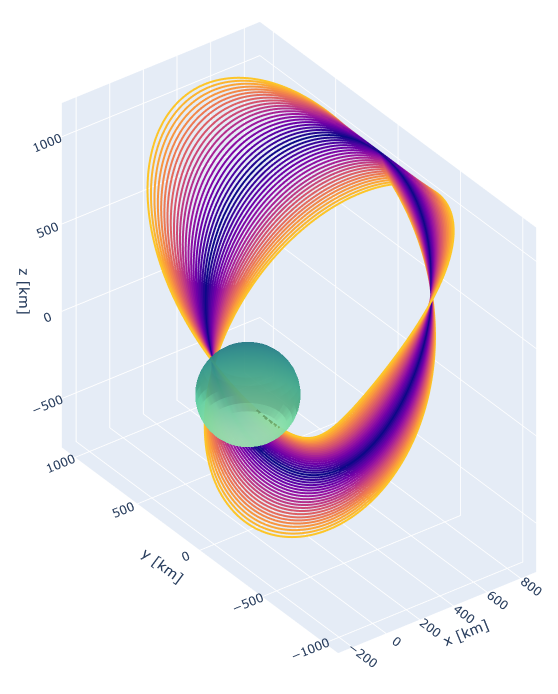} 
        \caption{Family of orbits generated from first period-doubling bifurcation of $L_2$ halo orbit family}
        \label{f:L2_halo_period_doubling_1}
    
        \centering
        \includegraphics[scale=0.6]{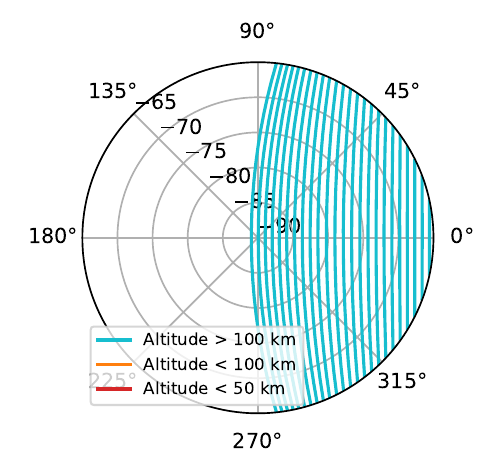} 
        \caption{Groundtracks over South pole for orbits generated from first period-doubling bifurcation of $L_2$ halo orbit family}
        \label{f:L2_halo_orbit_pd1_gts}

    \end{figure}

\begin{figure}

        \centering
        \includegraphics[scale=0.5]{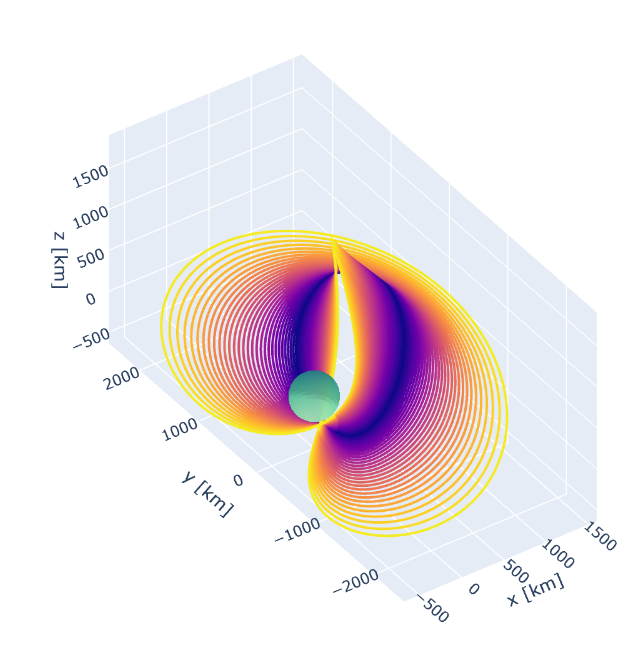} 
        \caption{Family of orbits generated from second period-doubling bifurcation of $L_2$ halo orbit family}
        \label{f:L2_halo_period_doubling_2}
    
        \centering
        \includegraphics[scale=0.6]{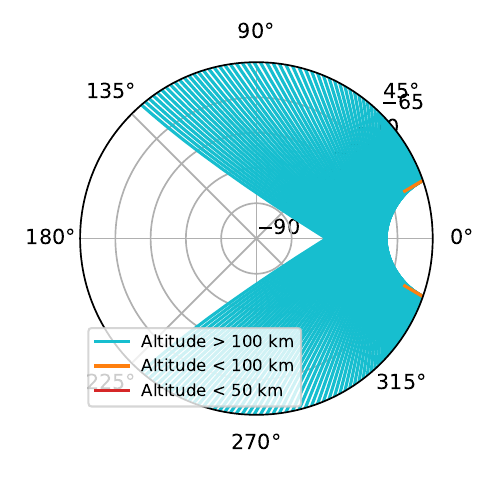} 
        \caption{Groundtracks over South pole for orbits generated from second period-doubling bifurcation of $L_2$ halo orbit family}
        \label{f:L2_halo_orbit_pd2_gts}
    
\end{figure}

The families obtained from the first two period-doubling bifurcations of the $L_2$ halo orbit family are shown in Figs.~\ref{f:L2_halo_period_doubling_1} and ~\ref{f:L2_halo_period_doubling_2}, and the family obtained from the period-tripling bifurcation is shown in Fig.~\ref{f:L2_halo_period_tripling}. All members of the family of orbits generated from the third period-doubling bifurcation were found to impact the surface of Enceladus, so these are not shown. The groundtracks for these families of orbits are shown in Figs.~\ref{f:L2_halo_orbit_pd1_gts},~\ref{f:L2_halo_orbit_pd2_gts} and~\ref{f:L2_halo_orbit_pt_gts}. These groundtracks show that none of these families contain orbits with close periapse passages over the South pole region, but do provide variation in groundtrack geometry with respect to the halo orbit families. 

\begin{figure}

        \centering
        \includegraphics[scale=0.5]{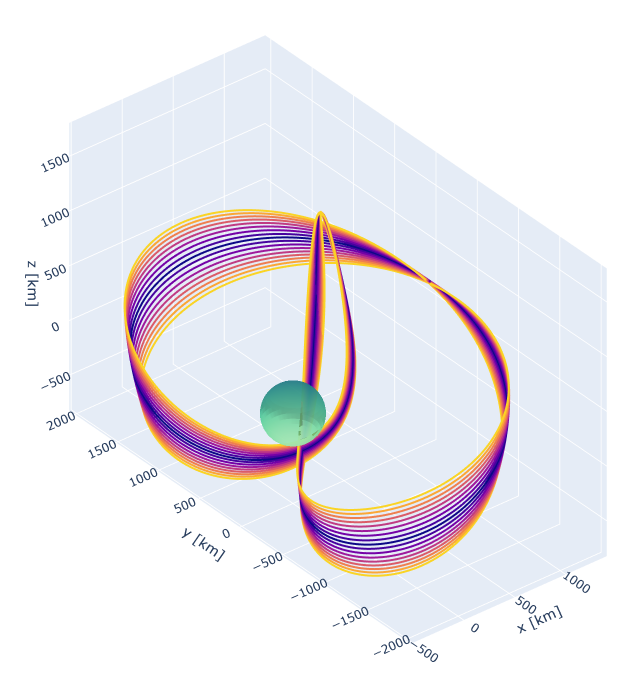} 
        \caption{Family of orbits generated from period-doubling bifurcation of family from Fig.~\ref{f:L2_halo_period_doubling_2}}
        \label{f:L2_halo_period_doubling_redoubling}
    
        \centering
        \includegraphics[scale=0.6]{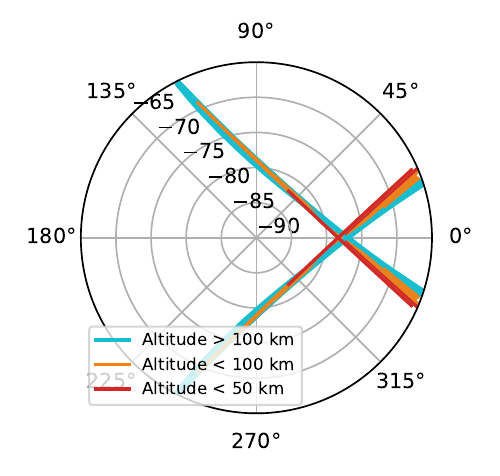} 
        \caption{Groundtracks over South pole for orbits generated from period-doubling bifurcation of family from Fig.~\ref{f:L2_halo_period_doubling_2}}
        \label{f:L2_halo_period_doubling_redoubling_gts}
    
\end{figure}

Further period-doubling and tripling bifurcations can subsequently be computed from these new families. An example of one such family is shown in Fig.~\ref{f:L2_halo_period_doubling_redoubling}. This family of orbits is obtained from a period-doubling bifurcation starting from the family shown in Fig.~\ref{f:L2_halo_period_doubling_2}. As seen in Fig.~\ref{f:L2_halo_period_doubling_redoubling_gts}, the resulting orbits have much closer periapse passes over the South pole region and could provide complementary coverage characteristics to the halo orbit family. We note that these more complex orbits may be more operationally challenging for navigation or stationkeeping purposes than the well-studied halo orbits, so further analysis would be needed to fully assess their suitability.

\begin{figure}
    
        \centering
        \includegraphics[scale=0.6]{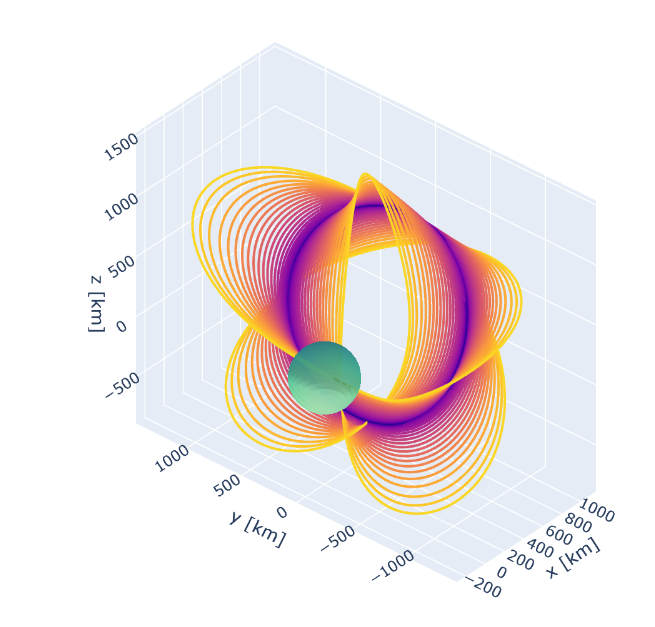} 
        \caption{Family of orbits generated from period-tripling bifurcation of $L_2$ halo orbit family}
        \label{f:L2_halo_period_tripling}
    
        \centering
        \includegraphics[scale=0.6]{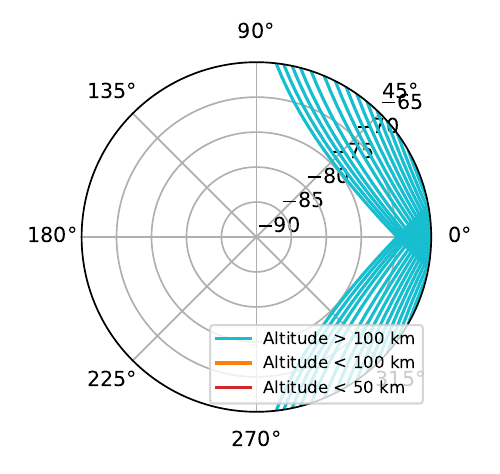} 
        \caption{Groundtracks over South pole for orbits generated from period-tripling bifurcation of $L_2$ halo orbit family}
        \label{f:L2_halo_orbit_pt_gts}

    \end{figure}

\subsection{Butterfly orbit family}
\begin{figure}
    \centering
    
        \centering
        \includegraphics[scale=0.5]{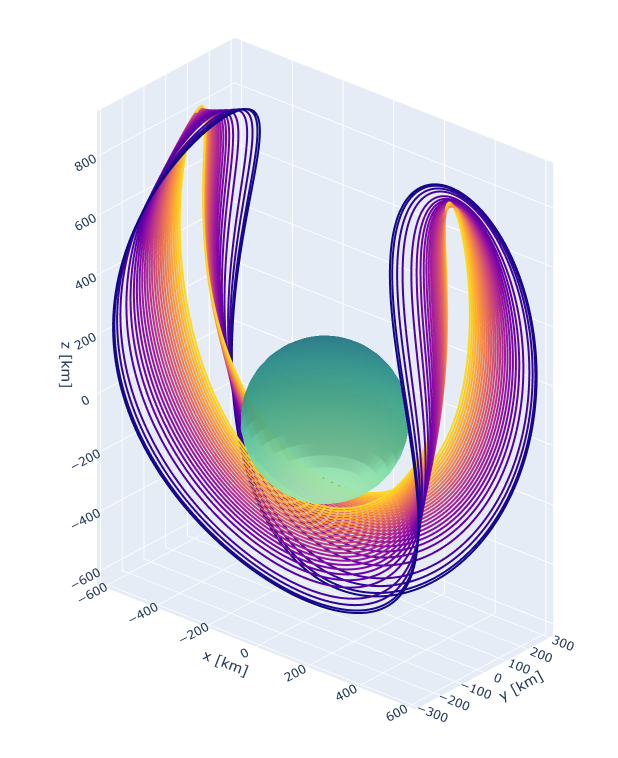} 
        \caption{Subset of butterfly orbit family}
        \label{f:butterfly_orbits}
    
        \centering
        \includegraphics[scale=0.6]{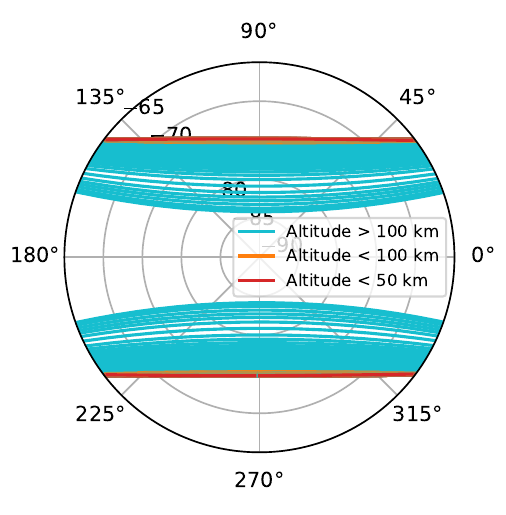} 
        \caption{Groundtracks over South pole for butterfly orbit family}
        \label{f:butterfly_gts}
        \includegraphics[scale=0.6]{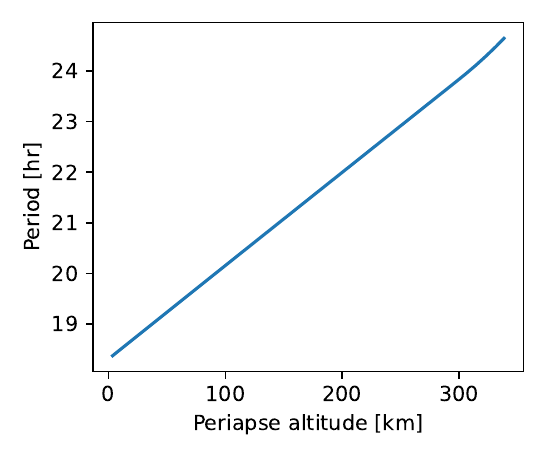} 
        \caption{Butterfly orbit periods vs. periapse altitude}
        \label{f:butterfly_period_alt}
    
\end{figure}

As stated previously, the period-doubling bifurcation from the halo orbit family resulting in the butterfly orbits does not appear in the Broucke diagram for the Saturn-Enceladus system. We can, however, continue these orbits from known initial conditions. We obtain initial conditions from the JPL Three-Body Periodic Orbit Catalog~\cite{jpl_database}, and continue the family of butterfly orbits using the SEMPy toolbox. A subset of this family is shown in Fig.~\ref{f:butterfly_orbits}. The groundtracks for these orbits are shown in Fig.~\ref{f:butterfly_gts}, showing that these orbits provide close passages over two different regions of Enceladus' South pole, which could provide complementary coverage to the halo orbit family.

\subsection{$L_4$ and $L_5$ axial orbit families}
The $L_4$ and $L_5$ axial orbit families bifurcate from the $L_4$ and $L_5$ vertical orbit families and can be continued until they reach the vicinity of the $L_1$ Lagrange point. In the Earth-Moon system, these families will eventually reach a bifurcation with the $L_1$ halo orbit family~\cite{doedel, grebow_thesis}; however, this bifurcation does not appear in the Saturn-Enceladus system. Example orbits from these families are shown in Fig.~\ref{f:axials}. 

\begin{figure}
    
        \centering
        \includegraphics[scale=0.4]{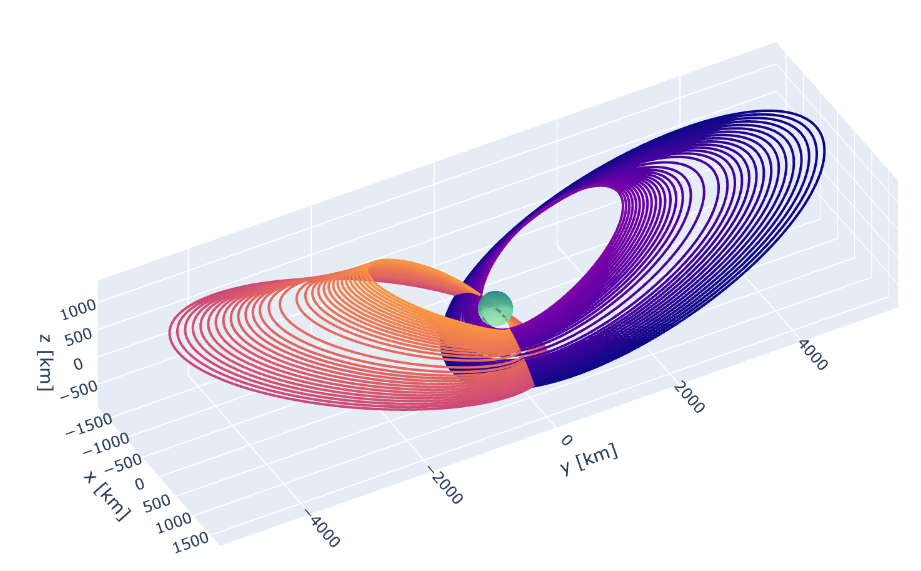} 
        \caption{Family of orbits generated from period-doubling bifurcation of $L_4$ and $L_5$ axial orbit families}
        \label{f:L2_halo_period_tripling}
\end{figure}

While the axial orbits themselves were not found to provide particularly useful geometries for repeated passes over the South pole region, bifurcations from these families could yield orbits with favorable and varying observation geometry over the South pole region. Examples of such families computed for the Jupiter-Callisto system can be found in Yang et al.~\cite{yang_families_jupiter_callisto}. Like the butterfly orbits, these families of periodic orbits will have groundtracks nearly perpendicular to those of the halo orbit family, and could be used in conjunction with the halo orbits to provide variability in the observation geometry.

\begin{figure}
    
        \centering
        \includegraphics[scale=0.35]{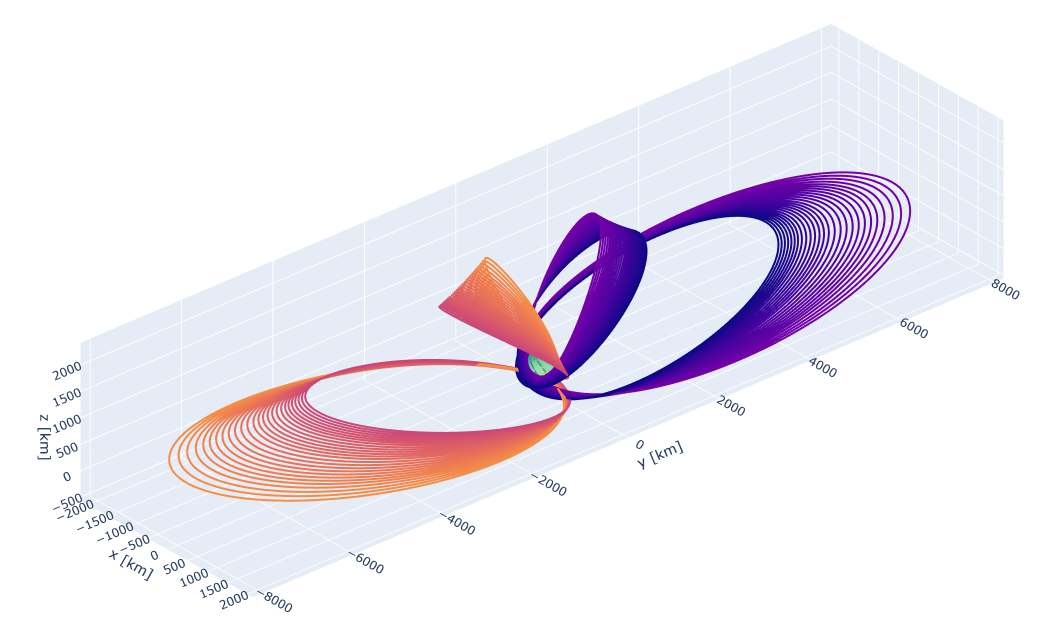} 
        \caption{Family of orbits generated from period-doubling bifurcation of $L_4$ and $L_5$ axial orbit families}
        \label{f:L2_halo_period_tripling}

        \centering
        \includegraphics[scale=0.6]{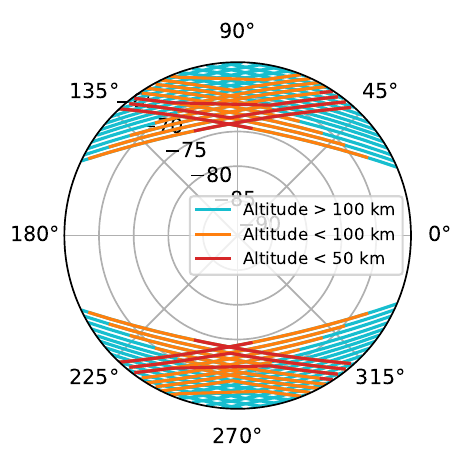} 
        \caption{Groundtracks over South pole for orbits generated from period-doubling bifurcation of $L_4$ and $L_5$ axial orbit families}
        \label{f:L2_halo_orbit_pt_gts}
    
\end{figure}

\section{Heteroclinic connection families}
As seen in Figs.~\ref{f:l1_halo_orbits} and~\ref{f:l2_halo_orbits}, the more distant halo orbits do not provide desirable observation geometries, but they remain relatively far away from the surface of Enceladus and could therefore provide safe ''loitering" zones for a spacecraft. As proposed in several works~\cite{salazar_cr3bp_oblate, salazar_low_energy, fantino_end_to_end_trajectory, davis_enceladus_nrho}, heteroclinic connections between these more distant halo orbits can provide low-energy transfers with potentially useful geometries requiring only small amounts of propellant. Heteroclinic connections are generated by connecting the unstable manifolds of the departure periodic orbit with the stable manifolds of the arrival periodic orbit. In Salazar et al.~\cite{salazar_cr3bp_oblate}, four types of heteroclinic connections were identified, comprising the various permutations of transfers between the Northern and Southern $L_1$ and $L_2$ halo orbits. The connections of each type with the lowest $\Delta V$ requirements were computed in the oblate CR3BP and analyzed.

 In this work we compute families of heteroclinic connections between distant halo orbits in the standard CR3BP. In order to compute these families, a simple continuation and targeting procedure was constructed. First, the SEMPy toolbox~\cite{sempy} is used to compute two halo orbits with the same Jacobi constant from the families we wish to connect. Next, the orbits are discretized according to mean anomaly, and the monodromy matrix for each orbit is computed at each mean anomaly. Then, the unstable manifold directions are computed for each location along the departure orbit, and the stable manifold directions are computed for each location along the arrival orbit. These are propagated forward or backward in time, respectively. A Poincare map~\cite{koon_hetconn} is generated by plotting the locations at which these manifolds intersect the $x=1-\mu$ plane. The two mean anomalies with the minimum position discontinuities in the $y$ and $z$ directions are chosen, and a single shooting differential corrections scheme is implemented to further reduce these position discontinuities. In this scheme, the mean anomalies are varied to reduce $\Delta y$ and $\Delta z$ until the norm of the two differences is below a certain tolerance, here set to $10^{-9}$. The velocity discontinuity between the two manifolds at the Poincare map intersection corresponds to the $\Delta V$ requirement for this particular connection. 
 
Starting from this first heteroclinic connection, a continuation scheme was developed to compute subsequent connections for halo orbits at different Jacobi constants. The differential corrections scheme can be implemented using the optimal mean anomalies from the previously computed connection as an initial guess, thus obviating the need to generate Poincare maps for each connection. Alternatively, to generate these families, the heteroclinic connections can be continued directly~\cite{haapala_hetconn}.

\begin{figure}
    \centering
    
        \centering
        \includegraphics[scale=0.55]{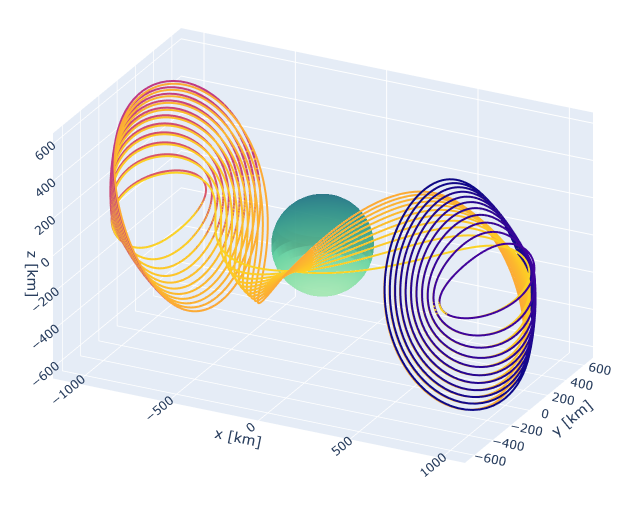} 
        \caption{Family of heteroclinic connections between Northern $L_1$ and Southern $L_2$ families}
        \label{f:type1_hetconn}
    
        \centering
        \includegraphics[scale=0.5]{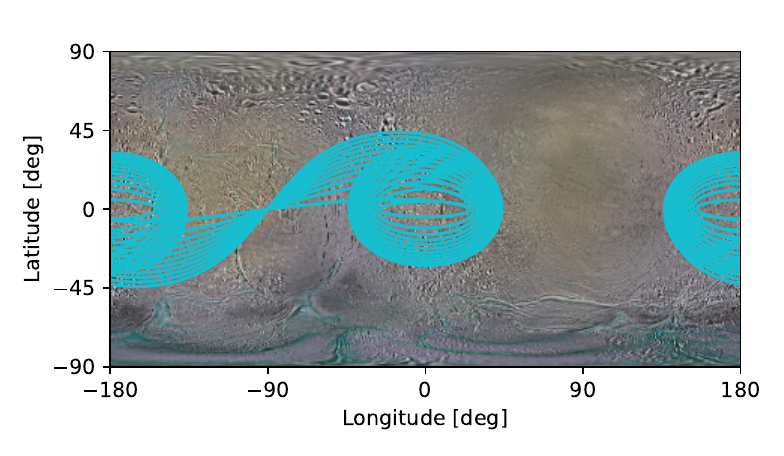} 
        \caption{Groundtracks for heteroclinic connections between Northern $L_1$ and Southern $L_2$ families}
        \label{f:type1_hetconn_gts}
        \includegraphics[scale=0.6]{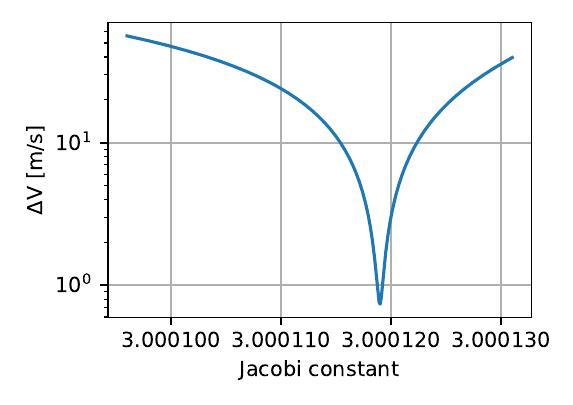} 
        \caption{$\Delta$V costs for heteroclinic connections between Northern $L_1$ and Southern $L_2$ families}
        \label{f:type1_hetconn_deltav}

\end{figure}

\begin{figure}
    
        \centering
        \includegraphics[scale=0.55]{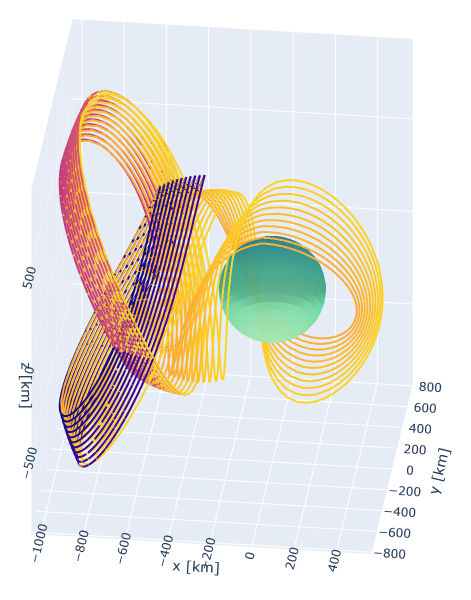} 
        \caption{Family of heteroclinic connections between Northern and Southern $L_1$ families}
        \label{f:type3_hetconn}
    
        \centering
        \includegraphics[scale=0.5]{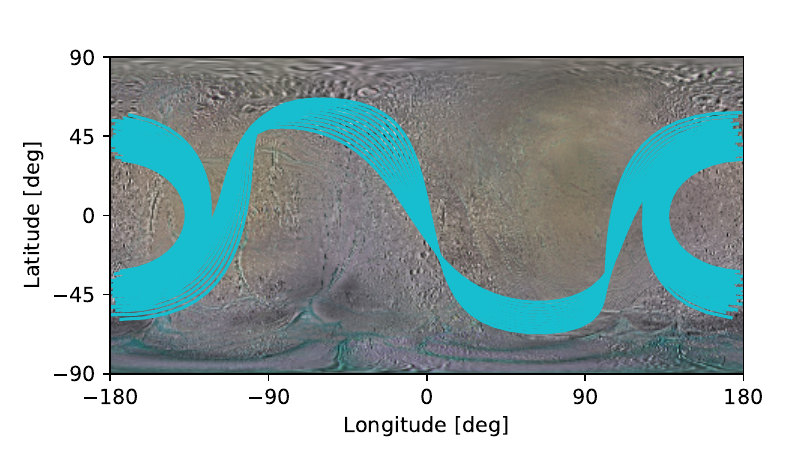} 
        \caption{Groundtracks for heteroclinic connections between Northern and Southern $L_1$ families}
        \label{f:type3_hetconn_gts}
        \includegraphics[scale=0.6]{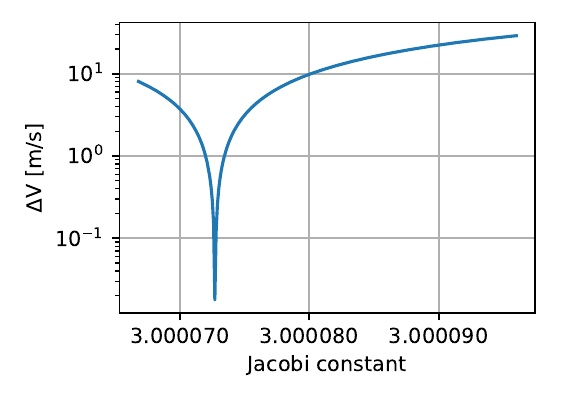} 
        \caption{$\Delta$V costs for heteroclinic connections between Northern and Southern $L_1$ families}
        \label{f:type3_hetconn_deltav}
    
\end{figure}

\begin{figure}
    
        \centering
        \includegraphics[scale=0.55]{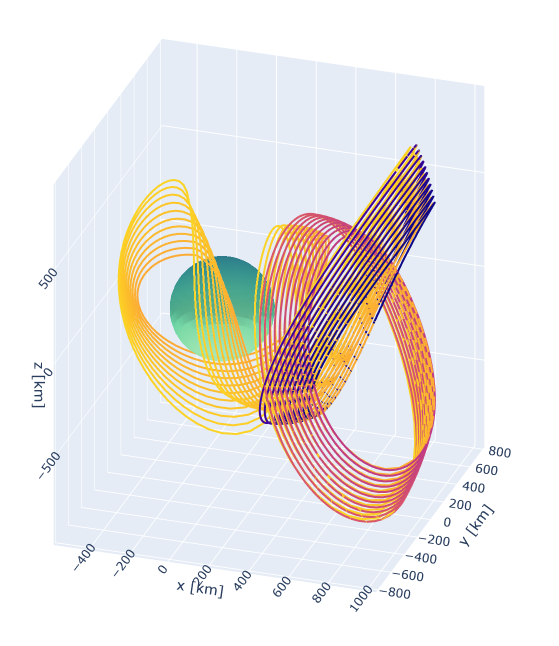} 
        \caption{Family of heteroclinic connections between Northern and Southern $L_2$ families}
        \label{f:type4_hetconn}
   
        \centering
        \includegraphics[scale=0.5]{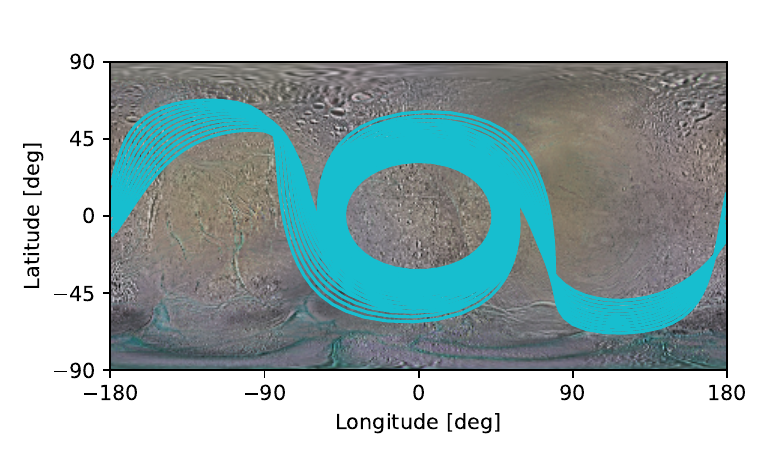} 
        \caption{Groundtracks for heteroclinic connections between Northern and Southern $L_2$ families}
        \label{f:type4_hetconn_gts}
        \includegraphics[scale=0.6]{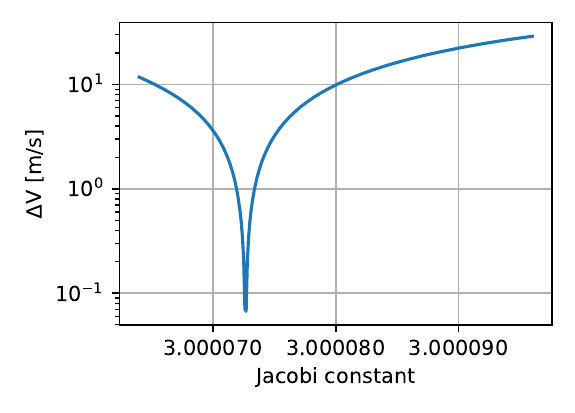} 
        \caption{$\Delta$V costs for heteroclinic connections between Northern and Southern $L_2$ families}
        \label{f:type4_hetconn_deltav}
    
\end{figure}

Heteroclinic connections are possible between the different combinations of Northern and Southern $L_1$ and $L_2$ halo orbit families. Families of connections between the Northern $L_1$ and Southern $L_2$ families are shown in Fig.~\ref{f:type1_hetconn}, between the Northern and Southern $L_1$ families in Fig.~\ref{f:type3_hetconn}, and between the Northern and Southern $L_2$ families in Fig.~\ref{f:type4_hetconn}. Groundtracks for each family of connections are shown in Figs.~\ref{f:type1_hetconn_gts},~\ref{f:type3_hetconn_gts} and~\ref{f:type4_hetconn_gts}. The $\Delta$V costs to link the unstable manifolds of the departure orbits with the stable manifolds of the arrival orbits are shown in Figs.~\ref{f:type1_hetconn_deltav},~\ref{f:type3_hetconn_deltav} and~\ref{f:type4_hetconn_deltav}. For each family of connections, there will be a pair of orbits for which the $\Delta$V costs for the connections will approach zero. The Jacobi constants found in this work for these orbits align with those found in previous works~\cite{salazar_cr3bp_oblate}. Nevertheless, the $\Delta$V costs to execute these low-energy transfers remain low for all connections in the families; as such, a variety of groundtrack geometries are available from these types of excursions. As seen in the groundtrack plots, these connections do not provide close passages over the South pole region, but they could be used to provide global mapping of Enceladus' surface without requiring insertion into a low-altitude orbit.

\section{Conclusions and future work}
In this work, we present an analysis and description of the dynamic structures available to use for science orbits configurations in the Saturn-Enceladus circular restricted three-body problem. These orbits can provide a variety of observation geometries for a future Enceladus-orbiting mission. In order to fully characterize these orbits, additional work is required to evaluate the operational feasibility, arrival costs, and station-keeping requirements for each type of orbit.

\bibliographystyle{AAS_publication}   
\bibliography{references}   

\end{document}